\documentstyle[12pt,epsfig]{article}
\def\bge{\begin{equation}}
\def\ene{\end{equation}}
\def\bg{\begin{eqnarray}}
\def\en{\end{eqnarray}}
\def\nn{\nonumber}

\begin{document}
\renewcommand{\thefootnote}{\fnsymbol{footnote}}
\begin{flushright}
ADP-97-33/T266 
\end{flushright}
%
%
\begin{center}
\begin{LARGE}
Quark and gluon condensates in the quark-meson coupling model 
\end{LARGE}
\end{center}
\vspace{0.5cm}
\begin{center}
\begin{large}
K.~Saito\footnote{ksaito@nucl.phys.tohoku.ac.jp} \\
Physics Division, Tohoku College of Pharmacy \\ Sendai 981, Japan \\
K.~Tsushima\footnote{ktsushim@physics.adelaide.edu.au} and 
A.~W.~Thomas\footnote{athomas@physics.adelaide.edu.au} \\
Department of Physics and Mathematical Physics \\
and \\
Special Research Center for the Subatomic Structure of Matter \\
University of Adelaide, South Australia, 5005, Australia
\end{large}
\end{center}
%
%
\begin{abstract}
Using the quark-meson coupling (QMC) model, we study the density dependence 
of the quark and gluon condensates in nuclear matter. 
We show that the change of the quark condensate is mainly driven by  
the scalar field in the medium and that the reduction of the quark 
condensate is suppressed at high density, even in the 
mean-field approximation.  The gluon condensate decreases by 4 - 6 \% at 
nuclear saturation density.  We also give a simple relationship between 
the change of the quark condensate and that of a hadron mass in the medium. 
\end{abstract}
\vspace{0.5cm}
\newpage

The QCD ground state is highly non-trivial, and the strong condensates of 
scalar quark-antiquark pairs $\langle {\bar q} q \rangle$ and gluon fields 
$\langle G_{\mu \nu}^a G^{a \mu \nu} \rangle$ may play important roles in 
a wide range of low-energy hadronic phenomena~\cite{shif,yaz,druk,hat}.  
Therefore, it is quite interesting to study the density dependence of the 
condensates in nuclear matter.  The vacuum values of the lowest-dimensional 
quark and gluon condensates are typically given by~\cite{yaz} 
\bge
Q_0 \equiv \langle {\bar q} q \rangle_0 \simeq - (225 \pm 25 \mbox{MeV})^3, 
\label{qcon0}
\ene
\bge
G_0 \equiv \langle G_{\mu \nu}^a G^{a \mu \nu} \rangle_0 \simeq (360 \pm 20 
\mbox{MeV})^4. 
\label{gcon0}
\ene

Drukarev et al.~\cite{druk}, Cohen et al.~\cite{cohen} and Lutz et 
al.~\cite{lutz} have shown that the leading 
dependence on the nuclear density, $\rho_B$, of 
the quark condensate in nuclear 
matter, $Q(\rho_B)$, is given by the model-independent form: 
\bge
\frac{Q(\rho_B)}{Q_0} \simeq 1 - \frac{\sigma_N}{f_\pi^2 m_\pi^2} \rho_B , 
\label{indep}
\ene
where $\sigma_N$ is the pion-nucleon sigma term (empirically $\sigma_N \simeq 
45$ MeV~\cite{gass}), $m_\pi$ is the pion mass (138 MeV) and $f_\pi \simeq 
93$ MeV, the pion decay constant.  
Further, the strange quark content in the nucleon at finite density (and 
temperature) was studied in Ref.~\cite{tsushima} using 
the Nambu--Jona-Lasinio (NJL) model, supplemented by an instanton 
induced interaction involving the in-medium quark condensates. 
The gluon condensate at finite density,  
$G(\rho_B)$, has also been discussed in Ref.~\cite{cohen}. 

Several years ago M. Ericson~\cite{ericson} suggested that a 
``distortion factor", coming 
from rescattering of soft pions in the nuclear medium, would tend reduce the 
amount of the chiral symmetry restoration (i.e., 
to oppose the reduction of the quark 
condensate).  However, Birse et al.~\cite{birse} 
pointed out that there was an 
incompleteness in the treatment of the rescattering of soft pions using 
the simple linear $\sigma$ model, and showed that the full amplitude for 
soft-pion scattering from two nucleons leads to an enhancement of the chiral 
symmetry restoration at finite density.  

Recently the in-medium quark condensate has been calculated in several, more 
elaborate ways~\cite{bhf,ways}.  In particular, using the 
Dirac-Brueckner-Hartree-Fock (DBHF) approach, Li and Ko and 
Brockmann and Weise 
have shown that higher-order contributions become increasingly 
important at high density, and tend to hinder the restoration of chiral 
symmetry~\cite{bhf}.  

We study the density dependence of the quark and gluon 
condensates in nuclear matter within the framework of the quark-meson 
coupling (QMC) model~\cite{gui,st,qmc1,qmc2,jen}.  
The QMC model may be viewed as an extension of QHD~\cite{qhd} in 
which the nucleons still interact through the exchange of scalar ($\sigma$) 
and vector ($\omega$ and $\rho$) mesons.  However, the mesons couple 
not to point-like nucleons but to
confined quarks (in the nucleon bag).  
In studies of infinite nuclear matter it 
was found that the extra degrees of freedom provided by the internal 
structure of the nucleon give an acceptable value for the incompressibility 
once the coupling constants are chosen to reproduce the correct saturation 
energy and density for symmetric nuclear matter. 
This is a significant improvement on QHD at the same level of 
sophistication (see also Ref.~\cite{nat}).  
Furthermore, the model has been successfully applied to 
finite nuclei within the Born-Oppenheimer 
approximation~\cite{qmc1,qmc2,blun,rho}. 
It has been found that the QMC model can reproduce the properties of 
finite, closed-shell nuclei (from $^{12}$C to $^{208}$Pb) quite well. 

As shown in Refs.~\cite{qmc1,qmc2}, the basic result in the QMC model with 
mean-field approximation (MFA) is that, in the scalar and vector meson 
fields, the nucleon behaves as if it had an 
effective mass $M_N^{\star}$. The latter can be 
calculated using a relativistic quark model 
for the nucleon (e.g., the MIT bag model) and depends 
on the nuclear density only through 
the $\sigma$ field. 

In an earlier version of the QMC model~\cite{qmc1,blun}, we considered 
the effect of the nuclear medium on the structure of the nucleon alone 
and froze the quark degrees of freedom in the mesons.  
We call this version QMC-I.  We have calculated the quark condensate in 
nuclear matter using this version~\cite{st,qcon0,qcon00}. 
(For a recent study, see also Ref.~\cite{mm}.) 
However, strictly speaking, those calculations (as well as the DBHF 
calculations~\cite{bhf}) were not complete because the meson structure 
effects were not treated consistently.  
The mesons themselves are built of quarks and anti-quarks, 
and their structure may also change in matter~\cite{qmc2,rho}.  
An additional, technical difference from Refs.~\cite{st,qcon0,qcon00} 
is that for reasons explained in Ref.~\cite{qmc1} -- see 
especially Appendix -- the c.m. correction to the bag energy is now 
treated as being independent of the applied scalar field.

To incorporate the effect of meson structure in the QMC model in MFA, we 
suppose that the vector mesons are again described by the MIT bag model 
with {\em common\/} scalar and vector mean-fields 
(like the nucleon in QMC-I).  In this case the effective vector-meson mass in 
matter, $m_v^{\star}$ $(v = \omega, \rho)$, will also depend on the $\sigma$ 
mean-field.  
The $\sigma$ meson itself is, however, not so readily represented  
by a simple quark model (like a bag), because it couples strongly 
to two pions and a direct 
treatment of chiral symmetry in medium is important~\cite{birse,birse2}.  
On the other hand, many approaches, including 
the NJL model~\cite{hat}, the Walecka 
model~\cite{sai} and Brown-Rho scaling~\cite{brown} suggest that the 
$\sigma$-meson mass in medium, $m_{\sigma}^{\star}$, should   
be less than the free value, $m_\sigma$ (= 550 MeV).  
It has therefore been parametrized as a quadratic 
function of the scalar field~\cite{qmc2}: 
\bge
\left( \frac{m_{\sigma}^{\star}}{m_{\sigma}} \right) = 1 - a_{\sigma} 
(g_{\sigma} {\bar \sigma}) + b_{\sigma} (g_{\sigma} {\bar \sigma})^2 , 
\label{sigmas}
\ene
with $g_\sigma {\bar \sigma}$ in MeV. Here 
${\bar \sigma}$ is the mean-field value of the $\sigma$ field and 
$g_\sigma$ is the $\sigma$-nucleon 
coupling constant (in free space). 
Three parameter sets: ($a_\sigma$ ; $b_\sigma$) = 
(3.0, 5.0 and 7.5 $\times 10^{-4}$ MeV$^{-1}$ ; 
10, 5 and 10 $\times 10^{-7}$ MeV$^{-2}$), called A, B and C, respectively,
were determined in Ref.~\cite{qmc2} 
so as to reduce the mass 
of the $\sigma$-meson by about 2\%, 7\% and 10\% (respectively) at 
saturation density $\rho_0$ $(= 0.15$ fm$^{-3})$.  This model, 
involving the structure 
effects of both the nucleon and the mesons, was called QMC-II~\cite{qmc2}.  

Within QMC-II, the total energy per nucleon, $E_{tot}$, can be written 
as~\cite{qmc2} 
\bge
E_{tot} = \frac{2}{\rho_B (2\pi)^3}\sum_{i=p,n}\int^{k_{F_i}} 
d\vec{k} \sqrt{M_i^{\star 2} + \vec{k}^2} + \frac{m_{\sigma}^{\star 2}}
{2\rho_B}{\bar \sigma}^2 + \frac{g_{\omega}^2}
{2m_{\omega}^{\star 2}}\rho_B + \frac{g_{\rho}^2}{8m_{\rho}^{\star 2}
\rho_B} \rho_3^2 , \label{tote}
\ene
where $g_v$ ($v= \omega, \rho$) is the $v$-nucleon coupling constant and 
$k_{F_i}$ ($i =$ proton or neutron) is the Fermi momentum.  The density of 
protons (neutrons) $\rho_p$ ($\rho_n$) is defined by 
$\rho_i = k_{F_i}^3/(3\pi^2)$, and then $\rho_B = \rho_p + \rho_n$ and 
$\rho_3 = \rho_p - \rho_n$.  Detailed values of the coupling constants and 
properties of nuclear matter in QMC-II can be found in Ref.~\cite{qmc2}.  

\begin{figure}[htb]
\begin{center}
\epsfig{file=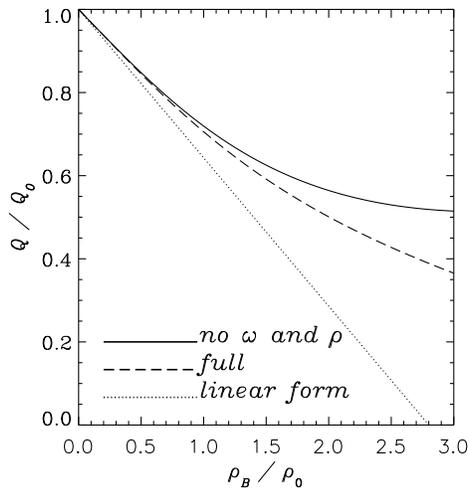,height=7cm}
\caption{Quark condensate at finite density using parameter set B.  
The dashed and dotted curves 
are respectively for the full calculation in symmetric nuclear matter 
($f_p=0.5$) and the linear approximation, Eq.(\protect\ref{qcon3}).  
The solid curve is the result corrected by removing the spurious 
$\omega$ and $\rho$ contributions.
}
\label{qcon}
\end{center}
\end{figure}

The density-dependent quark condensate, $Q(\rho_B)$, is formally derived by 
applying the Hellmann--Feynman theorem to the chiral-symmetry-breaking quark 
mass term of the total Hamiltonian.  One finds the relation for the 
quark condensate in nuclear matter at the baryon density $\rho_B$: 
\bge
m_q (Q(\rho_B) - Q_0) =  m_q \frac{d}{dm_q}  {\cal E}(\rho_B)  , 
\label{rel}
\ene
where ${\cal E}(\rho_B) = \rho_B E_{tot}$ and $m_q$ is 
the average, current quark mass of the u and d quarks.
The resulting value for $Q/Q_0$ as a function of density is 
shown in Fig.~\ref{qcon} -- dashed line. (We have chosen the quark mass
to be $m_q = 5$ MeV and the bag radius of the free nucleon 
$R_N = 0.8$ fm, but the result is quite insensitive to these choices.)

Using Eq.(\ref{rel}), the Gell-Mann--Oakes--Renner relation~\cite{gel} 
and the explicit expression for the self-consistency condition
of the $\sigma$ field in nuclear matter~\cite{qmc2},
Eq.(\ref{tote}) leads to the following explicit relation for the 
ratio of $Q(\rho_B)$ to $Q_0$:  
\bg
\frac{Q(\rho_B)}{Q_0} = 1 &-& \left( \frac{\sigma_N}{m_\pi^2 f_\pi^2} 
\right)  \left( \frac{m_\sigma^{\star}}{g_\sigma} \right)^2 (g_\sigma 
{\bar \sigma}) \left[ 1 - \frac{1}{g_\sigma} \left( \frac{d g_\sigma^q}{d m_q} 
\right) (g_\sigma {\bar \sigma}) \right] \nn \\
&-& \left( \frac{\sigma_N \rho_0}{6S_N(0) m_\pi^2 f_\pi^2} \right) 
\rho_r^2 \left[ \frac{\rho_0}{m_\omega^{\star 2}} 
\left( \frac{d g_\omega^2}{d m_q} \right) + 
\frac{\rho_0}{4 m_\rho^{\star 2}} (2 f_p -1)^2 
\left( \frac{d g_\rho^2}{d m_q} \right) \right] , 
\label{qcon1}
\en
where $\rho_r = \rho_B/\rho_0$, $f_p = \rho_p/\rho_B$, $S_N(0)$ is the quark 
scalar charge of the free nucleon ($ = \int_{bag} d{\vec r} {\bar \psi}_q 
\psi_q$) and $g_\sigma^q = g_\sigma/(3S_N(0))$~\cite{qmc1,qmc2}.  
The vector-meson mass is calculated using the bag model, while the 
$\sigma$-meson mass is given by Eq.(\ref{sigmas}). 
Because $m_q$ enters only in the combination $m_q - g^q_\sigma
\bar{\sigma}$, which is generally regarded as the
chiral-symmetry-breaking term in nuclear medium, we were able to
evaluate $\left( \frac{d m_\sigma^\star}{d m_q} \right)$ in terms of 
the derivative of $m_\sigma^\star$ with respect to the applied
scalar field ${\bar \sigma}$.

%

In Eq.(\ref{qcon1}) we have followed the usual convention of identifying 
$3 m_q S_N(0)$, which is the sigma commutator in the free 
MIT bag, as the experimental pion-nucleon sigma term, $\sigma_N$~\cite{st}. 
It is well known that the meson cloud of the nucleon (mainly the pions),
as well as its strange quark content contribute 
significantly to $\sigma_N$~\cite{jaffe}.  
However, because we are concerned primarily with the variation of $Q$ in matter 
from its free value, $Q_0$, it should be reasonable to replace $\sigma_N$ in 
Eq.(\ref{qcon1}) by its 
empirical value. (We note that the main
variation of $Q$ in medium is generated by the $\sigma$ mean-field.)

Clearly, from Eq.(\ref{qcon1}), the leading dependence of the quark 
condensate on the density is given by the scalar field:
\bge
\frac{Q(\rho_B)}{Q_0} \simeq 1 - \frac{\sigma_N}{m_\pi^2 f_\pi^2} 
\left( \frac{m_\sigma}{g_\sigma} \right)^2 (g_\sigma {\bar \sigma}) . 
\label{qcon2}
\ene
One can easily show that Eq.(\ref{qcon2}) reduces to the model-independent 
result, Eq.(\ref{indep}), to leading order in the density, so that 
for small $\rho_r$ one has (for the set B)~\cite{rho}:  
\bge
Q(\rho_B) / Q_0 \simeq 1 - 0.357 \rho_r . 
\label{qcon3}
\ene
This is shown as the dotted line in Fig.~\ref{qcon}.

Equation (\ref{qcon1}) also involves deviations of the 
quark-meson coupling constants
with respect to $m_q$. In principle, if one could derive these coupling 
constants from QCD, their dependence on $m_q$ would be given.
Within the present model there is no reason to believe that the couplings 
should vary with $m_q$. This is especially so for the vector couplings 
since they involve conserved vector currents. On the other hand, 
we require that our model reproduces the correct saturation energy 
and density of nuclear matter whatever parameters are chosen for the 
free nucleon. As a consequence, the coupling constants depend on 
$m_q$ in a way that has nothing to do with chiral symmetry breaking.
(For example, for set B, we find  
$g_\sigma^q = 4.891 - 0.005880  m_q + 1.200 \times 10^{-5} m_q^2, \ \ 
g_\omega^2 = 39.59 + 0.03828 m_q + 1.144 \times 10^{-3} m_q^2$ and 
$g_\rho^2 = 66.3 - 0.02 m_q$, with $m_q$ in MeV.)  

In order to extract a physically meaningful result for $Q/Q_0$ we 
should therefore remove the spurious contributions associated 
with $\frac{dg^q_M}{dm_q}$ 
$(M=\sigma,\rho,\omega)$ in Eq.(\ref{qcon1}).
In fact, the variation of $g^q_\sigma$ with $m_q$ is extremely small 
so we need only correct the $\omega$ and $\rho$ contributions. 
The final, corrected result is shown as the solid line in 
Fig.\ref{qcon}.
Even in the mean-field approximation, our calculations 
show that the higher-order contributions in the nuclear density become 
very important and that they weaken the chiral 
symmetry restoration at high density 
(c.f. Ref.~\cite{bhf}).
In QMC-II, the $\sigma$ field in nuclear matter 
is suppressed at high density (for example, $g_\sigma {\bar \sigma} \simeq$ 
200 (300) MeV at $\rho_0$ ($3 \rho_0$)) because the quark scalar charge, 
$S_N$, decreases significantly as the density rises, as a result of the 
change in the quark structure 
of the bound nucleon~\cite{qmc1,qmc2,st2}.  Since the reduction of the quark 
condensate is mainly controlled by the scalar field, it 
is much smaller than in the simple, linear approximation, 
Eq.(\ref{qcon3}).  
{}From the difference between the solid and dashed curves we see that the 
correction for the dependence of the coupling constants on $m_q$ 
is significant and this should be born in mind in any phenomenological 
treatment. 

We should note here that, from extensive studies of chiral perturbation 
theory for nuclear matter, especially the recent work of 
Birse~\cite{birse,birse2}, a reduction of the quark condensate from its 
vacuum value may {\em not\/} be enough to conclude that the chiral symmetry 
has been partially restored -- especially 
if part of the change in $\langle {\bar q}q 
\rangle$ arises from low-momentum pions.  We note also that higher-order 
condensates may play an increasingly important role as the quark condensate 
tends to zero.  

Next let us consider the in-medium gluon condensate.  
Cohen et al.~\cite{cohen} 
also developed a model-independent prediction of the gluon condensate that is 
valid to first order in the nuclear density through an application of the 
trace anomaly and the Hellmann--Feynman theorem.  

Following their approach, the 
ratio of the gluon condensate in nuclear matter, $G(\rho_B)$, to 
that in vacuum ($G_0$) is given by
\bge
G(\rho_B) / G_0 \simeq 1 - \left( \frac{8}{9 G_0} \right) 
\left[ {\cal E}(\rho_B) 
- 2 m_q (Q(\rho_B) - Q_0) - m_s (Q_s(\rho_B) - Q_{s0}) \right] , 
\label{gcon1}
\ene
where $m_s$ is the strange-quark mass and $Q_s(\rho_B)$ ($Q_{s0}$) is the 
strange-quark condensate in nuclear matter (in vacuum).  Up to first 
order in the density, the change of the strange-quark condensate may be 
written in terms of the strange quark content 
of the nucleon in free space, $S$: 
\bge
m_s (Q_s(\rho_B) - Q_{s0}) = S \rho_B + {\cal O}(\rho_B^2) . 
\label{stran}
\ene
The strange quark content is commonly specified by the dimensionless 
quantity, $y$, defined by 
\bge
y \equiv \frac{ 2 \langle {\bar s}s \rangle_N}
  {\langle {\bar u}u + {\bar d}d \rangle_N} , 
\label{yyy}
\ene
which leads to $S = (m_s/2m_q) \sigma_N y$.  
Roughly speaking, $y$ represents the probability to find $s$ or ${\bar s}$ in 
the nucleon and is a measure of the OZI-rule violation.  If 
$m_s/m_q \simeq 25$~\cite{gel} and $y \simeq 0.45$~\cite{cohen}, we get 
$S \simeq 250$ MeV.  We note, however, that $y \simeq 0.45$ is an 
extreme value, and it has recently been suggested that it may be 
compatible with zero~\cite{timm}.
In the analysis below we shall take care to examine the sensitivity 
to the full range of variation of $y$.

At very low $\rho_B$, ${\cal E}(\rho_B)$ can be expanded as~\cite{qhd} 
\bge
{\cal E}(\rho_B) = M_N \rho_B \left[ 1 + 
     \frac{3}{10 M_N^2} \left( \frac{3\pi^2}{2} \right)^{2/3} \rho_B^{2/3} 
     \right] + {\cal O}(\rho_B^2) ,
\label{eee}
\ene
where the second term in the bracket is the nonrelativistic Fermi-gas energy.  
Using the approximate form, 
$g_\sigma \bar{\sigma} \simeq 214$ (MeV) $\times \rho_r$  
(for the parameter set B in $m_\sigma^{\star}$)~\cite{rho},  
we find 
\bge
2 m_q (Q(\rho_B) - Q_0) = 214 (\mbox{MeV}) \times 
\sigma_N \left( \frac{m_\sigma}{g_\sigma} 
  \right)^2 \rho_r + {\cal O}(\rho_r^2). 
\label{qqq}
\ene
Choosing the central value of $G_0$ in Eq.(\ref{gcon0}), we then get the 
in-medium gluon condensate at low $\rho_B$ (for the set B): 
\bge
G(\rho_B)/G_0 = 1 - ( 0.03892 \rho_r + 0.001292 \rho_r^{5/3} ) + 
{\cal O}(\rho_r^2) . 
\label{gappr}
\ene
\begin{figure}[htb]
\begin{center}
\epsfig{file=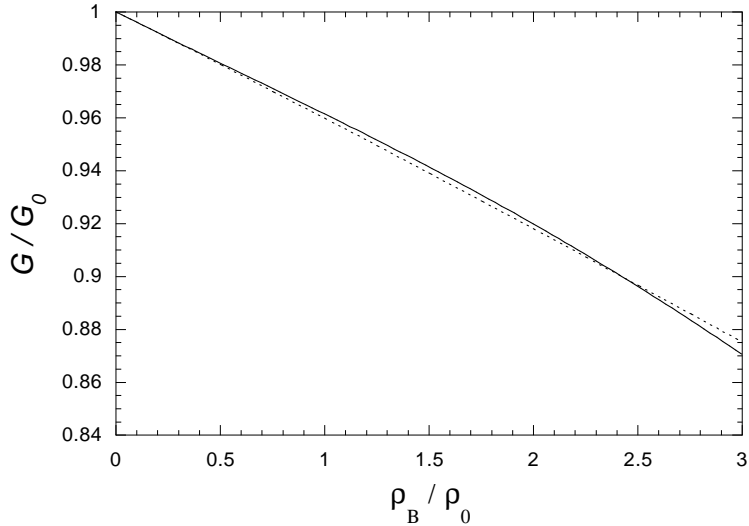,height=7cm}
\caption{Gluon condensate at finite density using parameter set B.  
The solid and dotted curves 
are respectively for the full calculation in symmetric nuclear matter 
($f_p=0.5$) and the approximation, Eq.(\protect\ref{gappr}). }  
\label{gcon}
\end{center}
\end{figure}
Our numerical results for the full calculation as well as the approximate 
calculation with Eq.(\ref{gappr}) are shown in Fig.~\ref{gcon} 
(for $m_q=5$ MeV and $R_N=$ 0.8 fm).
The reduction of the 
gluon condensate at finite density is not large; for example, it 
is reduced by only 4\% at $\rho_0$, 
which is consistent with the results of Cohen et al.~\cite{cohen}.  
The approximation of Eq.(\ref{gappr}) works very well for a wide range of the 
nuclear density, which may imply that the effect of 
higher-order contributions (in powers of the density)
is small for the gluon condensate.  However, one should keep in mind 
that the  
in-medium gluon condensate evaluated here 
contains a large uncertainty,
originating from the uncertainty in the value for the 
strange quark content of the nucleon in free space.  
We note that if we assume a 
vanishing strange quark content of the nucleon 
in free space ($S=0$ or $y=0$)~\cite{timm},  
the gluon condensate would be 
reduced by about 6\% at $\rho_0$.  

Finally, we relate the quark condensate to the variation of the hadron mass 
in nuclear matter.  In QMC-II, the hadron mass at low $\rho_B$ is simply given 
in terms of the scalar field~\cite{qmc2}: 
\bge
M_j^{\star} \simeq M_j - \frac{n_0}{3} (g_\sigma \sigma) , 
\label{efmas}
\ene
where $n_0$ is the number of non-strange quarks in the hadron $j$(= N, 
$\omega, \rho, \Lambda$, etc.).  Since the quark condensate at low $\rho_B$ is 
also determined by the scalar field  (see Eq.({\ref{qcon2})), we find a 
simple relation between the variations of the hadron mass and the quark 
condensate: 
\bge
\delta M_j^{\star} \simeq \left( \frac{m_\pi^2 f_\pi^2}{3 \sigma_N} \right) 
\left( \frac{g_\sigma}{m_\sigma} \right)^2 n_0 
\left( 1 - \frac{Q(\rho_B)}{Q_0} \right) \approx 200 (\mbox{MeV}) \times 
n_0 \left( 1 - \frac{Q(\rho_B)}{Q_0} \right) ,  
\label{relat}
\ene
where $\delta M_j^{\star} = M_j - M_j^{\star}$ (cf. Ref.~\cite{brown}).  

However, as shown in Ref.~\cite{birse2},  
we know that the nucleon mass in matter cannot depend in any simple way 
on the quark condensate alone because the leading non-analytic 
contribution (LNAC) to the pion-nucleon sigma term -- the term of 
order $m_\pi^3$ -- should {\em not\/} appear in the nucleon-nucleon 
interaction~\cite{ch}.  To discuss this problem further, we have to include 
pions 
self-consistently in the QMC model, which is beyond the 
scope of the present work.   

In summary, we have calculated the quark and gluon condensates in nuclear 
matter.  In the QMC-II model, the quark condensate at finite density is 
given in terms of the scalar field in the medium and the variation 
of the coupling constants with respect to the quark mass.  We have shown that 
the reduction of the quark condensate at high density is much less than 
that suggested by the 
(model-independent) leading-order prediction, Eq.(\ref{indep}), even in the 
mean-field approximation.
We also point out that the need to correct a naive use of the 
Hellmann--Feynman theorem to calculate $Q(\rho_B)$ for any purely 
phenomenological dependence of the quark-meson coupling constants 
on $m_q$. 

In comparison with the quark condensate, 
the gluon condensate does not decrease much in 
nuclear matter.  We have also provided a simple relationship  
between the change of the quark condensate and that of the hadron mass in 
nuclear matter.  We should notice here that the validity of our model 
is limited to low and moderate density (probably less than $\sim 3 \rho_0$),  
because the short-range correlations between 
quarks in overlapping hadron bags have been ignored.  The 
effect of the pion cloud of the hadrons~\cite{birse,birse2,tm} should 
also be considered explicitly in any 
truly quantitative study of the condensate properties in the medium.

\vspace{1cm}
We would like to thank M. Birse, P. Guichon and M. Ericson for helpful 
comments on the issues discussed here during the Workshop on 
Hadrons in Dense Matter held at the CSSM.
This work was supported by the Australian Research Council and  
the Japan Society for the Promotion of 
Science.
%
%

%
\end{document}